# Adaptive Mask-guided K-space Diffusion for Accelerated MRI Reconstruction

Qinrong Cai, Yu Guan, Zhibo Chen, Dong Liang, *Senior Member, IEEE,*
Qiuyun Fan, *Senior Member, IEEE* and Qiegen Liu, *Senior Member, IEEE*

*Abstract*—As the deep learning revolution marches on, masked modeling has emerged as a distinctive approach that involves predicting parts of the original data that are proportionally masked during training, and has demonstrated exceptional performance in multiple fields. Magnetic Resonance Imaging (MRI) reconstruction is a critical task in medical imaging that seeks to recover high-quality images from under-sampled k-space data. However, previous MRI reconstruction strategies usually optimized the entire image domain or k-space, without considering the importance of different frequency regions in the k-space This work introduces a diffusion model based on adaptive masks (AMDM), which utilizes the adaptive adjustment of frequency distribution based on k-space data to develop a hybrid masks mechanism that adapts to different k-space inputs. This enables the effective separation of high-frequency and low-frequency components, producing diverse frequency-specific representations. Additionally, the k-space frequency distribution informs the generation of adaptive masks, which, in turn, guide a closed-loop diffusion process. Experimental results verified the ability of this method to learn specific frequency information and thereby improved the quality of MRI reconstruction, providing a flexible framework for optimizing k-space data using masks in the future.

## I. INTRODUCTION

Magnetic Resonance Imaging (MRI) is an indispensable imaging technology in the medical field, and is widely used in clinical practice for the diagnosis, treatment and prevention of diseases. Due to the inherent characteristics of MR, which requires a long time to capture k-space data, there is an urgent demand for accelerated MRI strategies in the clinic to improve the imaging efficiency and facilitate subsequent diagnosis [1-4]. In response to this issue, a series of strategies have been proposed to under-sample the k-space data and then learn priors to reconstruct images. From sparsity-driven compressed sensing (CS) [5], [6] to the more recent deep learning-based models [7-9], reconstruction algorithms have made significant progress in clinical settings. However, due to the inherent complexity of the k-space frequency domain distribution, many existing methods face significant challenges in accurately modeling the underlying data distribution.

In response to the existing challenges, diffusion models (DMs) have demonstrated remarkable achievements in MRI reconstruction in recent years [10-15]. Among numerous related works, Jalal *et al.* [16] initially proposed the application of DMs in MRI reconstruction, pushing the boundaries of realistic MR data to reconstruct by applying posterior sampling with Langevin dynamics. Inspired by this innovative work, Song *et al.* [17] were committed to expanding the theoretical framework of DMs, who successfully extended them to medical image reconstruction tasks. Benefiting from the potent generative capabilities of DMs, certain studies have opted to face the challenges inherent in the k-space. Peng *et al.* exploited the redundancy and structural properties of the Hankel matrix to extract multiple low-rank patches from a single k-space data for prior learning [18]. During the iterative reconstruction process, they combined low-rank penalties and data consistency constraints to enhance the reconstruction quality. Tu *et al.* [19] trained a generative model by applying weighting techniques in the k-space, and achieved flexible calibration-free parallel imaging reconstruction by integrating traditional algorithms. However, the significant distribution differences between the high-frequency and low-frequency regions in the k-space make it still very difficult to capture the accurate k-space.

Given that mask modeling has gradually become a research hotspot in image reconstruction recently and has demonstrated excellent reconstruction performance in the context of natural images and other modalities, its application in MRI reconstruction also holds great potential [20-22]. Specifically, by introducing the mask mechanism, the model can capture the data distribution required by the task more accurately, thereby significantly improving the efficiency of prior learning. Based on this principle, Luo *et al.* performed region masking and reconstruction on medical images to learn useful representations [23]. Additionally, making the student model mimic the outputs of the teacher autoencoder via a hint loss, Wang *et al.* [24] computed the reconstruction loss for each patch and then selectively masked the patches that are more challenging to reconstruct. This approach ensures the model focuses on the most difficult

This work was supported in part by the National Key Research and Development Program of China under Grant 2023YFF1204300 and Grant 2023YFF1204302, in part by the National Natural Science Foundation of China under Grant 62122033. (Q. Cai and Y. Guan are co-first authors) (Corresponding author: Q. Liu)

Y. Guan is with School of Mathematics and Computer Sciences, Nanchang University, Nanchang 330031, China. (guanyu@email.ncu.edu.cn)

Q. Cai, Z Chen and Q. Liu are with School of Information Engineering, Nanchang University, Nanchang 330031, China. ({caiqinrong, 416100240222}@email.ncu.edu.cn, liuqiegen@ncu.edu.cn).

D. Liang is with the Lauterbur Research Center for Biomedical Imaging and the Research Center for Medical AI, Shenzhen Institute of Advanced Technology, Chinese Academy of Sciences, Shenzhen 518055, China (e-mail: dong.liang@siat.ac.cn).

Q. Fan is with the Academy of Medical Engineering and Translation-al Medicine, Medical School, Faculty of Medicine, Tianjin University, Tianjin 300072, China (e-mail: fanqiuyun@tju.edu.cn).

aspects of the data during training. The success of a large number of experiments has proved the feasibility of directly optimizing the k-space reconstruction [25-28]. It is worth noting that these schemes usually optimize the entire k-space while ignoring the importance of the distribution of different frequency components in the k-space [29-31]. Comprehensively considering the above factors, we embed the mask mechanism into the DMs, enabling the model to accurately learn the low-frequency contour and high-frequency detail information, which can accelerate and improve MRI reconstruction.

In this work, we propose an adaptive mask generation mechanism for MRI reconstruction called adaptive mask-based diffusion model (AMDM). Particularly, the proposed mask difference-based adaptive generation mechanism utilizes k-space characteristic by applying high-pass and low-pass filters to isolate frequency components, computing their differences with full-sampled k-space, and generating masks based on adaptive thresholds. This approach ensures precise frequency separation, enhancing reconstruction quality by emphasizing critical high-frequency details while maintaining structural integrity. Furthermore, hybrid masks strategy was adopted, enabling the model to solve the samples in multiple vector directions. This strategy enhances the learning from the frequency characteristics of k-space data, providing superior reconstruction quality and robustness. The main contributions of this work are summarized as follows:

● Unlike conventional masking strategies, our approach adaptively generates masks based on the input k-space data, allowing self-adjustment guided by the inherent frequency characteristics of the k-space. This ensures that the masks are dynamically tailored to the data, improving the accuracy of frequency-specific processing.

● Hybrid masks of different strengths are dynamically shaped according to the unique characteristics of each input k-space. This enables the DM to accurately estimate the upper bound of precision in the high-dimensional sample space formed by hybrid masks across multiple channels, thereby more effectively capturing the key features of different k-space data. The strategy enhances the flexibility and robustness of the model, thereby improving the reconstruction performance.

● K-space frequency distribution provides information for the generation of an adaptive mask, which in turn guides the diffusion process by filtering high- and low-frequency information in k-space. This process generates refined k-space data that iteratively influences the formation of the mask, forming a synergistic feedback loop. The closed-loop design ensures dynamic adjustment of information transfer, enabling the diffusion process to more accurately process signals of different frequencies, thus improving the accuracy of MRI reconstruction.

The structure of the manuscript is organized as follows: Section II provides a brief overview of related works. Section III presents the core concept of the proposed method. Section IV details the experimental setup and results. Section V offers a succinct discussion, and Section VI concludes the work.

## II. PRELIMINARIES

### A. Forward Imaging Model

MRI scanner collects signal in the k-space by the receiver coil, the forward model of MRI can usually be expressed as:

$$y = mk + \eta, \quad (1)$$

where $y \in \mathbb{C}^N$ represents the under-sampled measurement in the k-space, $k \in \mathbb{C}^N$ denotes the k-space to be reconstruct. $\eta$ is the Gaussian noise and $m$ is the under-sampled matrix. Due to the long scan duration of MRI reconstruction, many scans are accelerated by sampling k-space at a sub-Nyquist rate. In these cases, the forward model is often rank-deficient, making the recovery of the k-space is an ill posed problem. Further, the reconstruction of target image can be reformulated as an optimization problem:

$$k^* = \underset{k}{argmin}\{\|mk - y\|_2^2 + \mu R(k)\}, \quad (2)$$

where the first term is often called the data fidelity, which ensures that the reconstruction is consistent with the actual measurements. $R(k)$ is the prior constraint of the MR image.

### B. Score-based Diffusion Model

Score-based diffusion model [32] has made significant contributions across various generative domains [33-36]. Relying on its powerful generative capabilities, it can also flexibly adjust the diffusion behavior to accommodate different k-space data distributions. This adaptability allows it to better fit the data, thereby improving the accuracy and robustness of the reconstruction process. Typical model follows a two-stage process comprising that a forward diffusion process perturbing data to noise and a reverse diffusion process converting noise back to data. The forward diffusion process can be described by the following stochastic differential equation (SDE):

$$dx = f(x,t)dt + g(t)du, \quad (3)$$

where function $f$ is the drift function, $g$ is called the diffusion coefficient, and $u$ is the standard Wiener process. SDE is a commonly used strategy in medical imaging [37], [38]. The reverse process is also a diffusion process, modeled by the following reverse-time SDE:

$$dx = [f(x,t) - g(t)^2 \nabla_x \log p_t(x)]dt + g(t)d\bar{u}, \quad (4)$$

where $\bar{u}$ is a standard Wiener process when time flows backwards from $T$ to 0 [39], and $\nabla_x \log p_t(x)$ is known as the score function. Although the analytical form of $\nabla_x \log p_t(x)$ is generally intractable, it can be approximated by training a neural network $s_\theta(x,t)$ to estimate the score function $\nabla_x \log p_t(x)$ at each time step. Formally, we optimize the parameters $\theta$ of the score network with the following cost:

$$\theta^* = \underset{\theta}{argmin} \mathbb{E}_t\{\lambda(t) \mathbb{E}_{x_0} \mathbb{E}_{x_t|x_0}[\|s_\theta(x,t) - \nabla_x \log p_t(x_t|x_0)\|_2^2]\}, \quad (5)$$

where $\lambda(t)$ is a positive weighting function. We can generate $x_0$ from $x_T$ by solving this reversed SDE. Each unique pairing of these parameters influences the characteristics, behavior, and properties of the resulting SDE. For the well-recognized variation known as the Variance Exploding SDE (VE-SDE), the coefficients can be expressed as $f = 0$ and $g = \sqrt{d[\sigma^2(t)]/dt}$. Here, $\sigma(t)$ represents the noise scale that is again a monotonically increasing function.

### C. Masked Image Modeling

Masked Image Modeling (MIM), as a powerful technique of generative self-supervised learning, has shown great potential by reconstructing masked natural image regions for training [40], [41]. In view of the differences in the numerical distribution of frequencies in the k-space, mask modeling can mask the low-frequency or high-frequency regions, thereby reducing the complexity of the data and enabling the DMs to learn both local texture and global layout features more fully. The problem of MIM formally is defined as follows: An image $\mathbf{X} \in \mathbb{R}^{H \times W \times C}$ is partitioned into multiple patches $x \in \mathbb{R}^{N \times (P^2 C)}$, $x = [x]_{i=1}^{N}$ where $N$ denotes the number of patches. Masked sequence can be denoted as $x \odot \mathcal{M}$. The remaining unmasked patches $\widetilde{x}$ is used to reconstruct the original pixel through an encoder $f_\theta(\cdot)$ and a decoder $g_\theta(\cdot)$. We use $m_i$ to denote the hidden part at the masked portion as $m_i = f_\theta(\widetilde{x})$, The learning object is:

$$\mathcal{L}_{MIM} = \frac{1}{\|\mathcal{M}\|} \sum_{i \in \mathcal{N}} \mathrm{II}_{\{\mathcal{M}_i = 1\}} \mathcal{S}(m_i, x_i), \quad (6)$$

where $\mathrm{II}_{(\cdot)}$ is an indicator function, and $\mathcal{S}$ donates similarity measurement function. The reconstruction effect of the model is measured by calculating the error between the reconstructed mask patch $m_i$ and the original patch $x_i$. Based on this principle, the data that the model needs to process can be simplified, thus significantly improving the reconstruction efficiency [42-45]. Building on advances in MIM and DMs, Austin *et al.* [46] adopted mask diffusion and proposed a discrete diffusion formula for the first time. Similarly, Chen *et al.* [47] discussed the application of mask autoencoder in DMs, providing a theoretical basis for the combination of mask mechanism and model. In view of such a new diffusion framework, Shi *et al.* addressed complexity and accessibility challenges in existing masked diffusion models by developing flexible continuous time formulas [48]. By presenting a weighted integral of cross entropy loss, the optimization process is simplified.

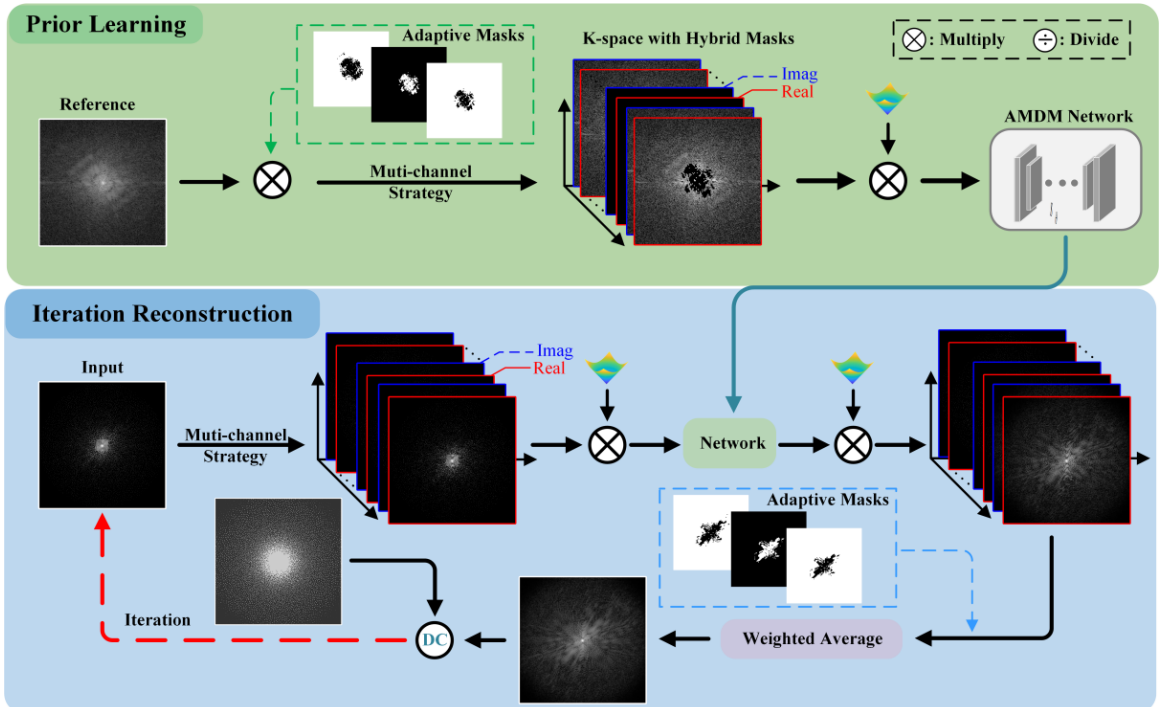

**Fig. 1.** Overall structure of AMDM. In the prior learning phase, multiple adaptive masks of different strengths and k-spaces first form a mask diffusion prior, and then form a high-dimensional tensor through a multi-channel strategy to enter the network for learning. During the iteration reconstruction, the dimension of the undersampled k-space data is changed through a multi-channel embedding strategy and iteratively updated in a high-dimensional data space. After the iteration is completed, the weights between channels are assigned according to the adaptive mask to enhance the sampling quality.

## III. METHODOLOGY

### A. Overall Structure of the Proposed AMDM

In MRI reconstruction, the learning of prior information mainly focuses on the complete k-space data. Although such method has achieved remarkable results, it ignores the importance of the distribution of different frequency components in the k-space. Meanwhile, relying solely on complete data may also limit the ability of model to capture local information, which is crucial for generating high-quality results. Therefore, we propose embedding the mask mechanism into the diffusion process and learning the priors from it. By utilizing the masked data to optimize the k-space data and thereby reduce its complexity, enabling the model to learn low-frequency contours and high-frequency detail information more accurately. This approach encourages the model to focus on the relation-ships between masked and unmasked regions, thereby enhancing its robustness and adaptability to various data distributions.

The overall structure of AMDM as shown in Fig. 1. Specifically, to realize the diversity on the form of modeling data, and more accurate prior information can be obtained, we introduce three innovations: Firstly, adaptive mask formulation is developed that dynamically adjusts the masking process based on the frequency characteristics of the input k-space, improving the ability of the model to capture local features. Secondly, data processed by dynamic hybrid masks via multi-channel strategy, enabling the model to accurately estimate the upper bound of accuracy in the high-dimensional sample space, thereby learning the fine features of the image more fully. It can be seen in Fig. 2 that the stacked distribution is more concentrated on each plane than the original data distribution, and the characteristics of the data become more prominent and easier to be captured.

Finally, closed-loop diffusion mechanism is introduced to integrate the MIM-based prior into the iterative reconstruction scheme. This forms a feedback loop where refined k-space data iteratively updates the mask, enabling dynamic adjustment and improving MRI reconstruction accuracy.

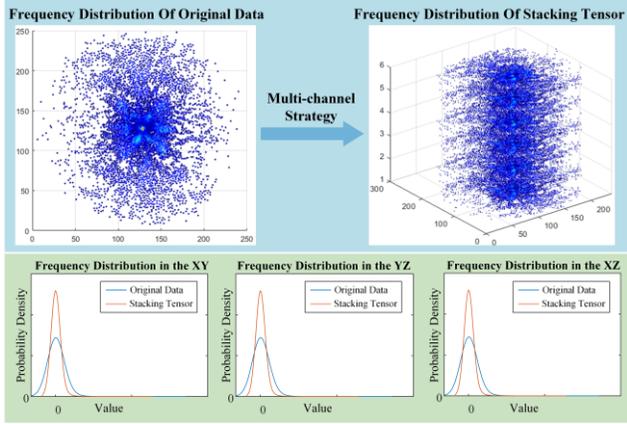

Fig. 2. Top: Frequency distribution of original data and stacking tensors. Bottom: Distribution comparison of original data and stacking tensor distribution on three axes.

### B. Adaptive Mask-Guided Priors for Diffusion Process

*Diffusion Prior with Adaptive Mask:* Dynamic mask adjustment tailored to the distinctive features of k-space data is achieved by designing filters $\mathbb{H}(\cdot)$ and $\mathbb{L}(\cdot)$. The high-pass filter $\mathbb{H}(\cdot)$ functions by initially decomposing the input signal via a wavelet transform into its constituent low-frequency and high-frequency components. Subsequently, it nullifies the low-frequency component, preserving solely the high-frequency components. Finally, the inverse transform is applied to these high-frequency components to reconstruct a signal. Correspondingly, the low-pass filter $\mathbb{L}(\cdot)$ removes the high frequencies by transforming them and then inverse transforms the remaining frequencies. By analyzing the value range of the high- and low-frequency components, threshold ranges $[\tau_{min}, \tau_{max}]$ and $[\varphi_{min}, \varphi_{max}]$ can be set, which can be manually adjusted to obtain multiple masks of different intensities as needed. The adaptive mask generation process is shown in Fig. 3.

Assume that the adaptive mask is a matrix $\mathcal{M} \in \{0,1\}^{N \times N}$, where $N$ is the size of the k-space data. The generation mechanism of the adaptive mask can be expressed as follows:

$$\begin{cases} \mathcal{M}_L = \mathbb{I}(\tau_{min} \leq |k - \mathbb{H}(k)| \leq \tau_{max}) \\ \mathcal{M}_H = \mathbb{I}(\varphi_{min} \leq |k - \mathbb{L}(k)| \leq \varphi_{max}) \end{cases}, \quad (7)$$

where $\mathbb{I}$ is the indicator function, which returns 1 if the condition is true and 0 otherwise. $k$ represents the k-space data. Based on the obtained adaptive masks, the frequency components can be correspondingly derived:

$$\begin{cases} k_H = k \odot \mathcal{M}_H \\ k_L = k \odot \mathcal{M}_L \end{cases}. \quad (8)$$

The adaptive mask is designed to capture the distinct frequency components of the k-space data, which are crucial for preserving fine details in the reconstructed images. During the training, the mask is updated to better match the distribution of the k-space, leading to improved reconstruction performance. However, a major problem with this approach is that it can lead to the loss of important information in the masked region, especially the low-frequency components, which are critical to maintaining the overall structure. To address these limitations, there are significant advantages to adopting a high-dimensional space embedding strategy. By embedding low-frequency components of k-space data into multiple high-frequency channels, the model is able to capture fine details and overall structure of the image, thereby improving the quality of reconstruction.

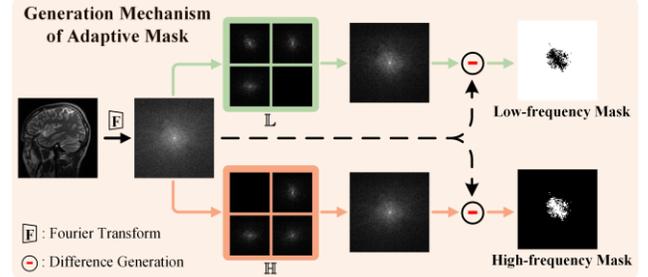

Fig. 3. Adaptive mask generation process. Input k-space is separated by high-pass filters $\mathbb{H}$ and low-pass filters $\mathbb{L}$. Differences from fully-sampled data are calculated, and adaptive masks are generated based on them and thresholds.

*High-dimensional Space with Hybrid Masks:* To ensure that critical information is not lost during the masked diffusion process and thus obtain a more accurate score function, we introduced a high-dimensional space embedding strategy to improve the reconstruction performance of masked modeling. More clearly, [49] describes error bounds for estimating the gradient of a data distribution in high-dimensional space and shows how this estimate can be optimized score function $\widetilde{s_\theta}$ by increasing the number of samples and the dimensionality of the space:

$$\widetilde{s_\theta} \in \underset{s_\theta \in D}{argmin} \mathbb{E}_t[||s_\theta(k,t) - \nabla_k \log p_t(k)||_2^2], \quad (9)$$

where $D$ is a class of $d$-dimensional functions, all of which are $M/2$-Lipschitz and restricted by $\mathcal{R} > 0$ at each coordinate, containing arbitrarily better approximations of $\nabla_k \log p_t(k)$ on the ball of radius $\mathcal{R}$. Then, with the randomness of the sample can obtain:

$$\mathbb{E}_t[||\widetilde{s_\theta} - \nabla_k \log p_t(k)||_2^2]$$
$$\leq C(M\mathcal{R} + \mathcal{B})^2(\log^3 n \cdot \mathcal{N}_n^2(F) + \beta_n d), \quad (10)$$

where $n$ represents the number of samples, $C$ is a universal constant. Both $\mathcal{N}_n^2(\cdot)$ and $\beta_n$ contain factor of $1/n$. It means that sampling in a high-dimensional space and increasing the number of embedding samples, especially the high frequency samples, can contribute to enhancing the performance of the model.

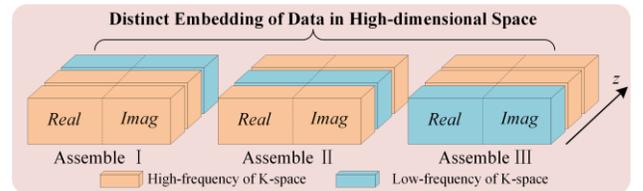

Fig. 4. Distinct embedding of frequency components in high-dimensional space.

Inspired by this high-dimensional space embedding strategy, we use multiple high-frequency components to build a high-dimensional tensor and embed low-frequency components at the same time to ensure that the model learns both more details and overall structural features of the image, thereby improving the robustness and generalization ability of the model. As shown in Fig. 4, according to the organization of channels in the high-dimensional tensor, the masked diffusion prior $\tilde{k}$ can achieve

more diverse and flexible combinations:

$$\tilde{k} = \begin{cases} \{k_L, k_{H_1}, \cdots, k_{H_N}\} \\ \{k_{H_1}, \cdots, k_L, \cdots, k_{H_N}\} \\ \{k_{H_1}, \cdots, k_{H_N}, k_L\} \end{cases} \quad (11)$$

Moreover, considering that the input MR images fall into complex-valued distributions and difficult for the score network to learn, we imitate [50] to process complex-valued data by connecting the real and imaginary parts as channels, that is, setting the real and imaginary parts as network inputs at the same time through the replication technique. Therefore, the masked diffusion prior can be expressed as:

$$\tilde{k} = \begin{cases} \{\mathcal{X}(k_L), \mathcal{X}(k_{H_1}), \cdots, \mathcal{X}(k_{H_N})\} \\ \{\mathcal{X}(k_{H_1}), \cdots, \mathcal{X}(k_L), \cdots, \mathcal{X}(k_{H_N})\}, \\ \{\mathcal{X}(k_{H_1}), \cdots, \mathcal{X}(k_{H_N}), \mathcal{X}(k_L)\} \end{cases} \quad (12)$$

where $\mathcal{X}(\cdot)$ is used to separate the real and imaginary parts of the data, e.g., $\mathcal{X}(k_L) = [(k_L)_{real}, (k_L)_{imag}]$. The aim of stacking to form $\tilde{k}$ is to solve the objects of subsequent network learning in the high-dimensional space, thereby averting potential issues in the score estimation accuracy and the Langevin dynamics sampling. In this case, the proposed AMDM set $N$ to be 2 during the training phase, and model is eventually trained with the following parameters $\theta^*$ to estimate the analytic form of the $\nabla_{\tilde{k}} \log p_t(\tilde{k})$:

$$\theta^* = \underset{\theta}{argmin} \mathbb{E}_t\{\lambda(t) \mathbb{E}_{\tilde{k}_0} \mathbb{E}_{\tilde{k}_t|\tilde{k}_0}[||s_\theta(\tilde{k}, t) \\ - \nabla_{\tilde{k}} \log p_t(\tilde{k}_t|\tilde{k}_0)||_2^2]\}. \quad (13)$$

### C. Closed-loop Diffusion for K-space Reconstruction

The generation process of the DM is initiated from a generalized numerical solution obtained by inverting the SDEs. Different from the practice of only using a numerical SDE solver to generate samples, we use the predictor-corrector (PC) sampler for iterative reconstruction, with the goal of achieving convergence. This choice is based on the fact that the PC sampler provides an alternative and potentially more optimized approach, as demonstrated in [17].

In the predictor stage, an initial estimate of the recovered image for the next sampling step is formulated. The corrector stage then refines this estimate using the Markov Chain Monte Carlo (MCMC) method [51], thereby improving the accuracy and quality of the recovered image. Formally, the predictor step modifies the generate image as follows:

$$\tilde{k}^i = \tilde{k}^{i+1} + (\sigma_{t+1}^2 - \sigma_t^2) s_\theta(\tilde{k}^{i+1}, t) + \sqrt{\sigma_{t+1}^2 - \sigma_t^2} \mathcal{Z}, \quad (14)$$

where $\sigma_t$ represents a monotonically increasing function with respect to time $t$. The variable $\mathcal{Z} \sim N(0, \mathbf{I})$ follows a Gaussian distribution random noise. Thereafter, the following corrector step can be utilized to refine the estimated image as follows:

$$\tilde{k}^i = \tilde{k}^i + \varepsilon_t s_\theta(\tilde{k}^i, t) + \sqrt{2\varepsilon_t} \mathcal{Z}, \quad (15)$$

Motivated by [15], the cascading of multiple score networks hold the potential to effectively enhance the quality of MRI reconstruction. Thus, based on the diversity of high-dimensional space embedding forms, we trained two different types of models $\mathfrak{D}_1$ and $\mathfrak{D}_2$ for iterative reconstruction. The overall iterative reconstruction structure can be expressed by formula:

$$\begin{cases} \tilde{k}^{i+1} = \{\mathcal{X}(k_{H_1}^{i+1}), \mathcal{X}(k_L^{i+1}), \mathcal{X}(k_{H_2}^{i+1})\} \\ \tilde{k}^i = \mathfrak{D}_1(\tilde{k}^{i+1}) = \{\mathcal{X}(k_{H_1}^i), \mathcal{X}(k_L^i), \mathcal{X}(k_{H_2}^i)\} \\ \tilde{k}^{i-1} = \mathfrak{D}_2(\tilde{k}^i) = \{\mathcal{X}(k_{H_1}^{i-1}), \mathcal{X}(k_L^{i-1}), \mathcal{X}(k_{H_2}^{i-1})\} \end{cases} \quad (16)$$

It should be emphasized that in each iteration of the annealed Langevin dynamics, the objects constructed by the hybrid masks must be mapped back from the high-dimensional space to the low-dimensional data space to enforce consistency constraints. Concurrently, the mask is dynamically updated to readapt to the evolving k-space frequency distribution generated in the previous iteration. This updated mask is then employed to guide the subsequent iterative reconstruction process. Let $\hbar$ represent the data from an arbitrary channel in $\tilde{k}$ and $\mathcal{M}$ is the corresponding adaptive mask, the return formula is as follows:

$$\begin{cases} \hbar_1^i = \mathfrak{D}_1(\hbar^{i+1} \odot \mathcal{M}_1) \\ \hbar^i = \hbar^{i+1} + \mathcal{M}_1^*(\hbar_1^i - \hbar^{i+1} \odot \mathcal{M}_1) \\ \hbar_2^i = \mathfrak{D}_2(\hbar^i \odot \mathcal{M}_2) \\ \hbar^{i-1} = \hbar_1^i + \mathcal{M}_2^*(\hbar_2^i - \hbar^i \odot \mathcal{M}_2) \end{cases}, \quad (17)$$

where the superscript $*$ is the Hermitian transpose operation. To leverage the robust generative capabilities of the model while ensuring consistency with the acquired data, a data consistency (DC) operation is incorporated following each generation step. Let $k^{i-1} = Mean(\tilde{k}^{i-1})$, the subproblem related to DC can be described as follows:

$$\underset{k}{min} \{\|mk - y\|_2^2 + \mu \|k - k^{i-1}\|_2^2\}. \quad (18)$$

Such a closed-loop diffusion strategy enables the model to focus on salient features within the k-space data, reconstructing fine details and structural information and reducing computational complexity by efficiently targeting relevant data features. **Algorithm 1** compactly represents the prior learning and iterative reconstruction process of AMDM.

---

**Algorithm 1: AMDM**

**Prior Learning Stage**

1. **Input:** K-space dataset via hybrid adaptive mask
   $k_{H_1} = k \odot \mathcal{M}_{H_1}, k_{H_2} = k \odot \mathcal{M}_{H_2}, k_L = k \odot \mathcal{M}_L$
2. $\tilde{k} = \tilde{k}_1$ or $\tilde{k}_2$ for the train processes:
   $\tilde{k}_1 = \{\mathcal{X}(k_{H_1}), \mathcal{X}(k_L), \mathcal{X}(k_{H_2})\}$
   $\tilde{k}_2 = \{\mathcal{X}(k_L), \mathcal{X}(k_{H_1}), \mathcal{X}(k_{H_2})\}$
3. **Output:** Learned score models $s_{\theta_1}; s_{\theta_2}$

**Iterative Reconstruction Stage**

1: **Setting:** $\tilde{k}^{i+1}, \sigma, \mathcal{Z} \sim N(0, \mathbf{I}), \varepsilon_t, T, \mathcal{M}_L, \mathcal{M}_H, \mu$
2: **For** $i = T$ **to** 1 **do**
3:   Execute score models $s_{\theta_1}$:
4:   $\tilde{k}^i = \tilde{k}^{i+1} + (\sigma_{t+1}^2 - \sigma_t^2) s_{\theta_1}(\tilde{k}^{i+1}, t) + \sqrt{\sigma_{t+1}^2 - \sigma_t^2} \mathcal{Z}$
5:   $\tilde{k}^i = \tilde{k}^i + \varepsilon_t s_{\theta_1}(\tilde{k}^i, t) + \sqrt{2\varepsilon_t} \mathcal{Z}$
6:   **Obtain** $\tilde{k}^i$ **by updating Eq. (17)**
7:   Execute score models $s_{\theta_2}$:
8:   $\tilde{k}^{i-1} = \tilde{k}^i + (\sigma_{t+1}^2 - \sigma_t^2) s_{\theta_2}(\tilde{k}^i, t) + \sqrt{\sigma_{t+1}^2 - \sigma_t^2} \mathcal{Z}$
9:   $\tilde{k}^{i-1} = \tilde{k}^{i-1} + \varepsilon_t s_{\theta_2}(\tilde{k}^{i-1}, t) + \sqrt{2\varepsilon_t} \mathcal{Z}$
10:  **Obtain** $\tilde{k}^{i-1}$ **by updating Eq. (17)**
11: **End for**
12: Update $\tilde{k}^{i-1}$
13: $k^{i-1} = Mean(\tilde{k}^{i-1})$
14: **Data Consistency:**
    $\underset{k}{min} \{\|Ak - y\|_2^2 + \mu \|k - k^{i-1}\|_2^2\}$
15: **Return** $k$

## IV. EXPERIMENTS

### A. Experimental Setup

*Datasets:* We selected the brain dataset **SIAT** for training, which is provided by Shenzhen Institute of Advanced Technology, the Chinese Academy of Sciences and informed consent of the imaging subject is obtained in accordance with Institutional Review Board policy. 500 complex-valued images of them are selected as the training data which are collected from healthy volunteers using a T2-weighted turbo spin echo sequence on a 3.0T scanner. These fully-sampled data are acquired by a 12-channel head coil with matrix size of 256×256 and combined to single-channel complex-valued data. To train more fully and enhance model performance., the dataset is expanded to 4000 sample through flip and rotation.

During the iterative reconstruction stage, we tested the reconstruction performance of model on in-vivo datasets with different sequences to verify its generalization ability. Specifically, **T1-GE Brain** are MR images including 8 coils complex-valued obtained using 3.0T GE. The FOV is $220 \times 220 mm^2$, and TR/TE is $17.6/11ms$. **T1-weighted Brain** are obtained from healthy volunteers with a T1-weighted, 3D spoiled gradient echo sequence in a 1.5T MRI scanner using joint-only coil. The FOV is $20 \times 20 \times 20 cm^3$, TR/TE is $17.6/8ms$ and flip angle is $20°$. Moreover, the *fastMRI+* dataset [52] was utilized to validate the clinical performance as well. For the *fastMRI+* public brain data, we preprocessed the target image size to $256 \times 256$ for reconstruction using the model trained on the **SIAT** brain dataset. Note that **Test1** and **Test2** are images with rich texture details in the *fastMRI+* dataset which were used for comparison with other k-space generative models.

*Parameter Configuration:* The parameter selection and training pipeline for DM are established following the recommendations in [17]. To effectively capture prior information, the noise variance schedule in the forward diffusion process is configured with $\sigma_{max} = 378$ and $\sigma_{min} = 0.01$. The batch size is set to 2 and the noise scale was set to $N = 1000$. Network optimization is performed using the Adam optimizer, with momentum parameters set to $\beta_1 = 0.9$ and $\beta_2 = 0.999$. To maintain data consistency after a single Langevin sampling step, we adopt the internal circulation with $M = 1$, ensuring high-quality reconstruction. Additional parameters are adjusted empirically within the suggested ranges to optimize model performance. Training and testing are conducted on dual NVIDIA 2080 Ti GPUs (12 GB). To guarantee a consistent evaluation of the proposed approach, the sampling process parameters remain unchanged across all datasets. Moreover, since the MRI data processed by the adaptive mask mechanism is complex-valued, it is divided into real and virtual components before being fed into the network, with each component acting as a separate channel for the multi-channel real data input.

*Performance Evaluation:* To quantitatively evaluate the reconstruction quality of the various reconstruction methods, we use metrics such as Peak Signal-to-Noise Ratio (PSNR), Structural Similarity Index (SSIM), and Mean Squared Error (MSE). The results of each method are adjusted to the best aiming to illustrate the advantages of AMDM. Note that the larger PSNR and SSIM, the smaller MSE, indicate a better reconstruction. Associated with the work of open-source code is available in: *https://github.com/yqx7150/AMDM.*

TABLE I
PSNR, SSIM, AND MSE (*E⁻⁴) COMPARISON WITH STATE-OF-THE-ART METHODS UNDER DIFFERENT SAMPLING WITH VARYING ACCELERATION FACTORS

| T1-GE Brain | | | | | | |
|---|---|---|---|---|---|---|
| Pattern | AF | SAKE | ESPIRiT | EBMRec | E2E-Varnet | AMDM |
| Poisson | R=8 | 37.31/0.9063/1.248 | 36.72/0.9044/2.129 | 36.61/0.9431/2.182 | 37.23/0.9433/1.356 | **41.93/0.9582/0.641** |
| | R=10 | 36.12/0.8923/1.553 | 36.02/0.9106/2.501 | 34.33/0.9177/3.692 | 35.06/0.9332/3.119 | **41.03/0.9485/0.789** |
| | R=12 | 35.04/0.8802/2.138 | 34.87/0.8944/3.256 | 32.45/0.8845/5.691 | 33.22/0.9223/4.761 | **39.98/0.9367/1.004** |
| | R=15 | 33.60/0.8649/4.112 | 32.29/0.8812/5.907 | 30.20/0.8287/9.561 | 31.93/0.9067/6.408 | **38.50/0.9291/1.411** |
| T1-weighted Brain | | | | | | |
| Pattern | AF | SAKE | ESPIRiT | EBMRec | E2E-Varnet | AMDM |
| Random | R=8 | 29.87/0.6851/19.964 | 33.92/0.7651/4.058 | 31.38/0.7056/7.278 | 32.68/0.7693/5.389 | **38.34/0.8271/1.466** |
| | R=10 | 28.99/0.6774/27.047 | 30.14/0.7443/9.688 | 30.39/0.6865/9.140 | 30.40/0.7407/8.475 | **36.90/0.8184/2.042** |
| | R=12 | 27.85/0.6672/38.767 | 29.77/0.7287/10.551 | 29.08/0.6675/12.367 | 30.07/0.7395/9.837 | **36.30/0.8144/2.342** |
| | R=15 | 27.12/0.6598/57.462 | 28.56/0.7111/13.946 | 28.63/0.6248/13.715 | 29.12/0.7239/12.239 | **34.01/0.7869/3.975** |

### B. Reconstruction Experiments

*Comparisons with State-of-the-arts:* To evaluate the efficacy of the proposed method, we performed comprehensive comparisons against conventional techniques SAKE [53] and ESPIRiT [54], as well as deep learning-based approaches EBMRec [55] and E2E-Varnet [56]. Table I presents the average PSNR, SSIM, and MSE values for dataset reconstruction with different acceleration factor and pattern. Quantitative analysis reveals that AMDM method consistently surpasses existing methods across all metrics. Notably, AMDM demonstrate remarkable performance even under Poisson and Random sampling with an acceleration factor of ×15, as evidenced by its relatively high PSNR and SSIM reconstruction metrics.

Visual assessments of the quantitative results as shown in Fig. 5-6. Obviously, traditional methods demonstrate critical limitations SAKE exhibits severe noise and aliasing artifacts, whereas ESPIRiT suffers from spatial blurring and structural loss. While deep learning methods show improved performance, EBMRec introduces texture distortion and localized artifacts under high acceleration factors, and E2E-Varnet excessively smooths high-frequency details. In contrast, AMDM achieves optimal noise suppression and detail fidelity. The visual enlarged area and error maps in Fig. 5-6 visually confirm the superiority of AMDM in structural preservation and artifact reduction.

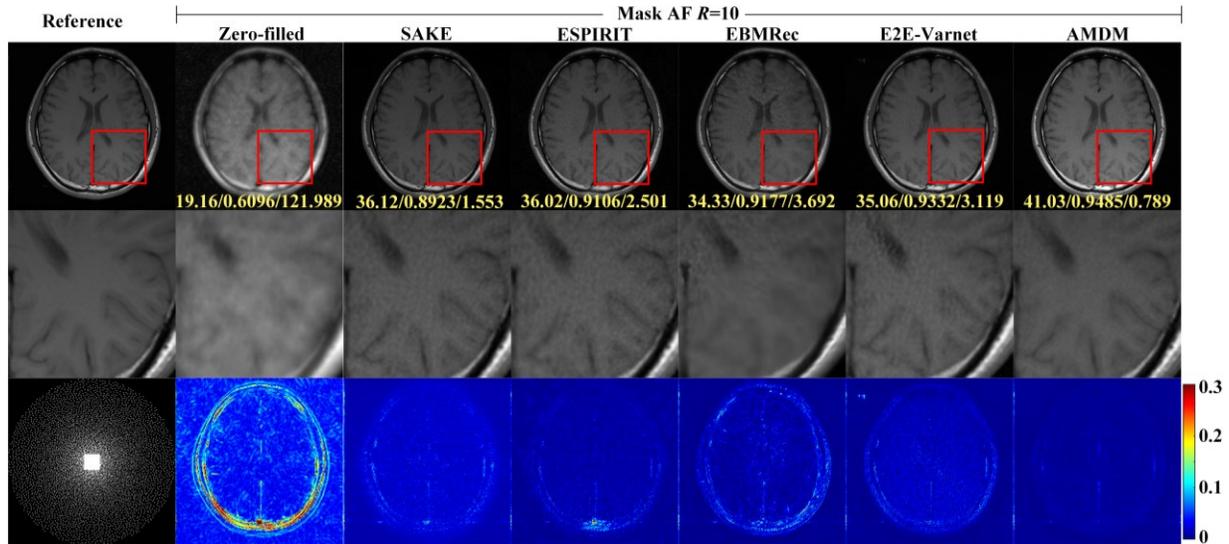

**Fig. 5.** Reconstruction of the *T1-GE Brain* at Poisson sampling of *R*=10. From left to right: zero-filled, reconstruction by SAKE, ESPIRIT, EBMRec, E2E-Varnet, AMDM and the reference. The second row shows the enlarged view of the ROI region (indicated by the yellow box in the first row), and the third row shows the error map of the reconstruction.

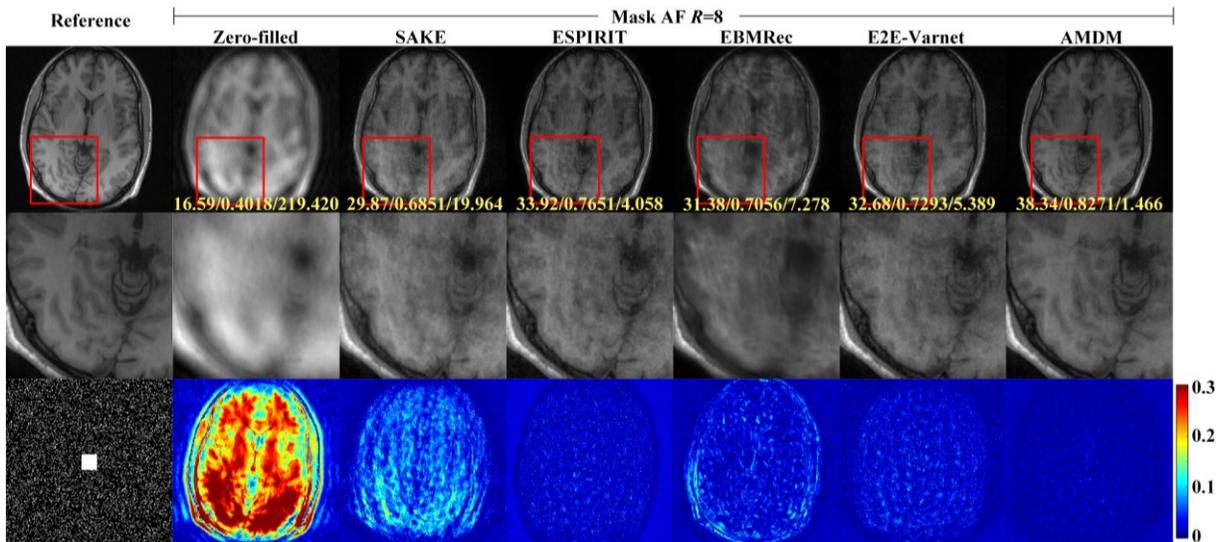

**Fig. 6.** Results of the *T1-weighted Brain* at *R*=8 Random sampling, with methods arranged sequentially from left to right: zero-filled, SAKE, ESPIRIT, EBMRec, E2E-Varnet, AMDM, and the reference. The second row presents enlarged ROI views (indicated by the yellow box in the first row), while the third row displays reconstruction error maps.

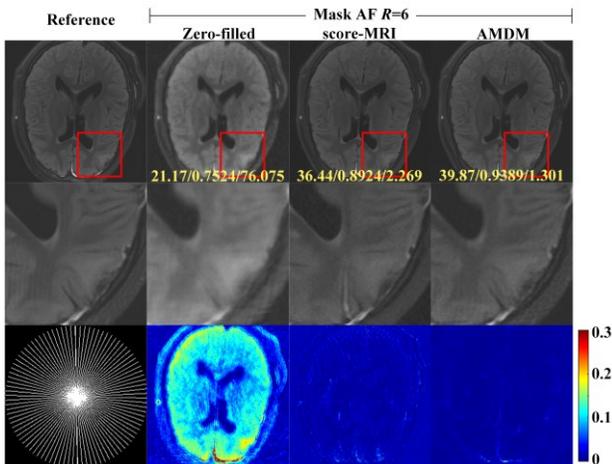

**Fig. 7.** Reconstruction results of score-MRI and AMDM under Radial sampling at *R*=6. Second row illustrates the error map.

TABLE II
QUANTITATIVE EVALUATIONS ON TEST 1 IN *FASTMRI+* DATASET WITH DIFFERENT ACCELERATION FACTORS

| *Test1* | | Zero-filled | score-MRI | AMDM |
|---|---|---|---|---|
| | ¶ | 21.17 | 36.44 | **39.87** |
| Radial *R*=6 | § | 0.7524 | 0.8924 | **0.9389** |
| | ‡ | 76.075 | 2.269 | **1.031** |
| | ¶ | 16.52 | 35.32 | **35.93** |
| Radial *R*=10 | § | 0.5819 | 0.8710 | **0.8893** |
| | ‡ | 223.291 | 2.935 | **2.551** |

¶: PSNR (dB) ↑, §: SSIM ↑, ‡: MSE (*$10^{-4}$) ↓

*Advantages with Masked Diffusion:* To demonstrate the advantages of masked diffusion, a primary quantitative comparison was conducted between our proposed AMDM, and the typical DM, score-MRI [57]. The average quantitative metrics for *Test 1* in the *fastMRI+* dataset under different acceleration factors and Poisson sampling are summarized in Table II. The results reaffirm the superiority of masked diffusion in terms of PSNR, SSIM, and MSE. Specifically, AMDM significantly outperforms score-MRI in the reconstruction of highly under-sampled images, showcasing its enhanced stability and reconstruction quality even at high acceleration factors. Visual results presented in Fig. 7 further illustrate the impact of masked diffusion. Under a Radial sampling pattern at an acceleration factor of $R$=6, AMDM demonstrates a clear advantage in preserving high-frequency details and reducing artifacts compared to score-MRI. The improved performance can be attributed to the masked diffusion process, which allows for more effective learning of lower-dimensional score functions, leading to refined training coverage and enhanced reconstruction quality.

TABLE III
PSNR, SSIM, AND MSE (*10$^{-4}$) MEASURES FOR DIFFERENT ALGORITHMS WITH VARYING ACCELERATE FACTORS

| *Test2* | | Zero-filled | WKGM | AMDM |
|---|---|---|---|---|
| Poisson *R*=8 | ¶ | 20.14 | 38.98 | **40.08** |
| | § | 0.7486 | 0.9169 | **0.9298** |
| | ‡ | 97.792 | 1.264 | **0.983** |
| Poisson *R*=12 | ¶ | 18.74 | 36.58 | **38.07** |
| | § | 0.7010 | 0.8927 | **0.9081** |
| | ‡ | 134.146 | 2.197 | **1.560** |

¶: PSNR (dB) ↑, §: SSIM ↑, ‡: MSE (*10$^{-4}$) ↓

*Influence with Adaptive Mask:* To highlight the benefits of adaptive masking, we conducted a comparative analysis between our proposed AMDM method and the WKGM [19]. As shown in Table III, AMDM consistently outperforms WKGM across all cases. Moreover, as the acceleration factor increases, WKGM exhibits rising reconstruction uncertainty, with residual noise hindering the preservation of structural details, as observed in Fig. 8. In contrast, AMDM effectively identifies and corrects over-smoothing and distortion, reconstructing accurate texture details with minimal noise. These results indicate that AMDM learns the distinct characteristics of both high and low-frequency priors, achieving high-quality reconstructions and demonstrating the clear advantages of adaptive mask in diffusion process.

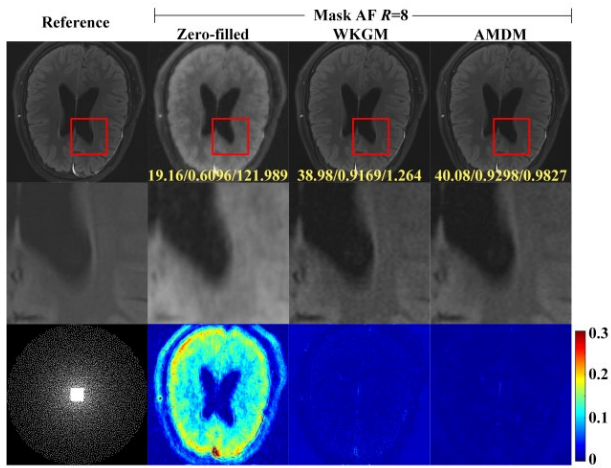

**Fig. 8.** Reconstruction results using WKGM and AMDM at *R*=8 of Poisson pattern. Second row displays error views.

### C. Convergence Analysis and Computational Cost

*Convergence Analysis:* In this subsection, we explore the convergence of WKGM and AMDM in terms of PSNR and SSIM through quantitative experiments. Specifically, we randomly selected an example of reconstructing brain images using a sampling pattern with an acceleration factor of *R*=8 to illustrates the correlation between convergence and the number of iterations. As shown in Fig. 9, with the increase of iterations, the PSNR and SSIM curves of both WKGM and AMDM rose rapidly at first, and then gradually converged. Ultimately, in a limited 800-step iteration, our approach achieved the best metrics in both. In particular, AMDM gradually showed a convergence trend within 100 steps and reached its peak in fewer iterations to converge. Obviously, using the adaptive mask mechanism to guide the diffusion process can effectively promote the capture of high-frequency information and accelerate the convergence speed. In addition, the hybrid masks enable the model to estimate samples in high-dimensional space, so as to capture key features of different k-space data more effectively, thus improving the training coverage and model performance.

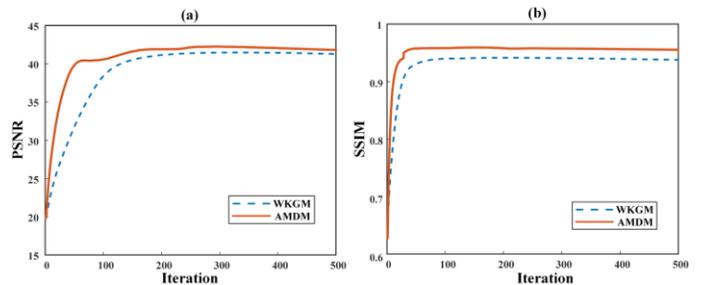

**Fig. 9.** Convergence curves of WKGM and AMDM in terms of PSNR (a) and SSIM (b) versus the iteration step from 1/8 sampled brain data.

*Practical Considerations on Diffusion Time:* Due to the obvious limitation of using DMs for image reconstruction is the time required for sampling, we further explored the diffusion time in Complete sampling and partial sampling method under actual experimental conditions, to prove the superiority of adaptive mask mechanism. According to the results in Table IV, it can be seen that the WKGM and AMDM have relatively few iterations and result in relatively lower time consumption. It means that the using the method of channel stacking can enable the model to estimate samples more accurately in a higher-dimensional space. Meanwhile, limited by the inherent characteristics of the DM in the entire space, the WKGM forward process occurs in the data distribution of the entire k-space, and its complex distribution further increases the computational cost and reduces the sampling speed. There is no doubt that the hybrid mask embedded with adaptive mask and high-dimensional channel strategy will reduce the running time, maximizing the advantages between them will further balance the reconstruction performance and the number of iterations. Intuitively, quantitative metrics also confirm that AMDM outperforms WKGM in terms of iteration time and ensures the quality of the reconstructed results.

TABLE IV
COMPUTATIONAL COST (UNIT: SECOND) OF DIFFERENT METHODS

| *Test 1* | score-MRI | WKGM | AMDM |
|---|---|---|---|
| Total-time (s) | 11244.66 | 1880.42 | **1308.54** |
| Iter-step | 861 | 334 | **193** |
| Iter-time (s) | 13.06 | 5.63 | **6.78** |

9## V. DISCUSSION

*Out-of-Distribution Performance:* In the preceding sections, we have comprehensively shown that the masked diffusion process goes beyond the reliance on superficial k-space data features and instead extracts essential information through adaptive mask strategy. By incorporating masked diffusion, a more robust latent prior is established, thereby significantly enhancing the generalization capacity of DMs. To substantiate this assertion, we performed experiments using the knee dataset with models trained exclusively on the brain dataset. The models evaluated include score-MRI, and AMDM. In this context, the test data distribution is distinct from the training data distribution, and the relevant experimental results are depicted in Table V. While score-MRI generally exhibits strong generative capabilities and achieve acceptable outcomes in out-of-distribution tasks, the proposed AMDM method capitalizes on its distinct advantages in masked diffusion to consistently produce superior reconstruction performance.

TABLE V
COMPARISON OF PSNR, SSIM AND MSE (*E$^{-4}$) OF DIFFERENT COMBINATION MANNERS IN POISSON SAMPLING PATTERNS

| Pattern | score-MRI | WKGM | AMDM |
|---|---|---|---|
| $R$=3 | 35.68/0.8652/2.702 | 36.82/0.9605/2.082 | **38.80/0.9629/1.317** |
| $R$=4 | 34.72/0.8385/3.377 | 34.86/0.9339/3.263 | **36.22/0.9411/2.386** |

*Marginal Effects of Frequency Complementarity:* We compare the reconstruction results using different channels to determine the optimal configuration. The purpose of channel superposition is to improve the quality of reconstruction by using the redundancy of multiple channels, that is, the larger the number of samples, the higher the spatial dimension and the smaller the estimation error [58]. However, increasing the number of channels may not always lead to proportional improvements in reconstruction quality due to the increased computational burden and potential overfitting. The metrics listed in Table VI demonstrate that the 6-channel configuration achieves the best reconstruction performance under Poisson sampling. One of the main reasons is that the 6-channel setup provides a balanced trade-off between redundancy utilization and computational efficiency. Additionally, the embedding of low-frequency reduces the risk of overfitting, allowing the model to generalize better across different datasets. In contrast, the 8-channel configurations, while offering more redundancy, suffer from increased computational complexity and potential overfitting, leading to slightly inferior reconstruction quality. This suggests that simply increasing the number of channels does not necessarily enhance reconstruction performance and may even degrade it due to the associated challenges.

TABLE VI
COMPARISON OF PSNR, SSIM AND MSE (*E$^{-4}$) OF DIFFERENT COMBINATION MANNERS IN 2D RANDOM SAMPLING PATTERNS

| | Channel Number | 4-*ch* | 6-*ch* | 8-*ch* |
|---|---|---|---|---|
| | PSNR | 40.36 | **41.93** | 40.82 |
| $R$=8 | SSIM | 0.9566 | **0.9582** | 0.9520 |
| | MSE | 0.920 | **0.641** | 0.827 |

## VI. CONCLUSIONS

In this work, we have proposed a novel MRI reconstruction method based on an adaptive mask-guided diffusion model. Our approach leverages the advantages of DMs and MIM to address the limitations of traditional methods that rely heavily on complete data modeling. Through the introduction of adaptive mask formulation and hybrid masks strategy, our model not only fully considers the differences in the distribution of frequency components in the k-space, but also optimizes the learning of complex priors. The adaptive mask formulation dynamically adjusts to the characteristics of the input data, improving the capture of local features. The hybrid masks strategy, combined with channel stacking, effectively handles complex data structures, refining the learning of priors. Additionally, our closed-loop diffusion mechanism efficiently integrates the MIM-based prior into the iterative reconstruction scheme, optimizing the utilization of prior information and improving reconstruction accuracy. Experimental results demonstrated that our method outperformed conventional approaches in terms of reconstruction quality and robustness. Future work will focus on further optimizing the model architecture and exploring its application in other medical imaging modalities.

## REFERENCES

[1] K. P. Pruessmann, M. Weiger, M. B. Scheidegger, et al., "SENSE: sensitivity encoding for fast MRI," *Magnetic Resonance in Medicine: An Official Journal of the International Society for Magnetic Resonance in Medicine*, vol. 42, no. 5, pp. 952–962, 1999.

[2] M. A. Griswold, P. M. Jakob, R. M. Heidemann, et al., "Generalized autocalibrating partially parallel acquisitions (GRAPPA)," *Magn. Reson. Med.*, vol. 47, no. 6, pp. 1202–1210, 2002.

[3] F. Knoll, C. Clason, K. Bredies, *et al.*, "Parallel imaging with nonlinear reconstruction using variational penalties," *Magn. Reson. Med.*, vol. 67, no. 1, pp. 34–41, 2011.

[4] J. Huang, P. F. Ferreira, L. Wang, et al., "Deep learning-based diffusion tensor cardiac magnetic resonance reconstruction: a comparison study," *Scientific Reports*, vol. 14, no. 1, pp. 5658, 2024.

[5] A. Majumdar, "Improving synthesis and analysis prior blind compressed sensing with low-rank constraints for dynamic MRI reconstruction," *Magn. Reson. Imag.*, vol. 33, no. 1, pp. 174–179, 2015.

[6] D. L. Donoho, "Compressed sensing," *IEEE Transactions on information theory*, vol. 52, no. 4, pp. 1289–1306, 2006.

[7] A. J. Sederman, L. F. Gladden, M. D. Mantle, "Application of magnetic resonance imaging techniques to particulate systems," *Advanced Powder Technology*, vol. 18, no. 1, pp. 23–38, 2007.

[8] R. Stannarius, "Magnetic resonance imaging of granular materials," *Review of Scientific Instruments*, vol. 88, no. 5, 2017.

[9] B. Zhu, J. Z. Liu, S. F. Cauley, et al., "Image reconstruction by domain-transform manifold learning," *Nature*, vol. 555, no. 7697, pp. 1476–4687, 2018.

[10] Z. Xiao, Y. Lu, B. He, et al., "Diffusion model based on generalized map for accelerated MRI," *NMR in Biomedicine*, vol. 37, no. 12, pp. e5232, 2024.

[11] M. Geng, J. Zhu, X. Zhu, et al., "DP-MDM: Detail-preserving mr reconstruction via multiple diffusion models," *Phys. Med. Biol.*, vol. 70, no. 11 pp. 115004, 2024.

[12] W. Zhang, Z. Xiao, H. Tao, et al., "Low-rank tensor assisted K-space generative model for parallel imaging reconstruction," *Magnetic Resonance Imaging*, vol. 103, pp. 198–207, 1999.

[13] C. Yu, Y. Guan, Z. Ke, et al., "Universal generative modeling in dual domains for dynamic MRI," *NMR in Biomedicine*, vol. 36, no. 12, pp. e5011, 2023.

[14] Y. Guan, Q. Cai, W. Li, et al., "Sub-DM: Subspace diffusion model with orthogonal decomposition for MRI reconstruction,". *arXiv preprint arXiv:*2411.03758, 2024.

[15] Y. Guan, C. Yu, Z. Cui, et al., "Correlated and multi-frequency diffusion modeling for highly under-sampled MRI reconstruction," *IEEE Trans. Med. Imag.*, vol. 43, no. 10, pp. 3490-3502, 2024.

[16] A. Jalal, M. Arvinte, G. Daras, et al., "Robust compressed sensing MRI with deep generative priors," *Advances in Neural Information Processing Systems*, vol. 34, pp. 14938-14954, 2021.